\documentclass[5p]{elsarticle}
\usepackage{amsmath,amssymb,graphics}

\newcommand{\hatA}[0]{\hat{A}}

\newcommand{\tr}[0]{\text{tr}}

\usepackage{grffile}
\begin{document}

\begin{frontmatter}
\title{The static quark potential from the gauge invariant Abelian decomposition}
\author[a]{Nigel Cundy}
\author[b,c]{Y. M. Cho}
\author[a]{Weonjong Lee}
\author[a]{Jaehoon Leem}
\address[a]{Lattice Gauge Theory Research Center, FPRD, and CTP  \\ 
Department of Physics and Astronomy, Seoul National University, 
Seoul, 151-747, South Korea}
\address[b]{Administration Building 310-4, Konkuk University,
Seoul 143-701, Korea}
\address[c]{Department of Physics \& Astronomy,
Seoul National University, Seoul, 151-747, Korea}

\begin{abstract}
We investigate the relationship between colour confinement and
topological structures derived from the gauge invariant Abelian
(Cho-Duan-Ge) decomposition.  This Abelian decomposition is made
imposing an isometry on a colour field $n$ which selects the Abelian
direction; the principle novelty of our study is that we have defined
this field in terms of the eigenvectors of the Wilson Loop. This
allows us to establish an equivalence between the path ordered
integral of the non-Abelian gauge fields with an integral over an
Abelian restricted gauge field which is tractable both theoretically
and numerically in lattice QCD. By using Stokes' theorem, we can
relate the Wilson Loop in terms of a surface integral over a
restricted field strength, and show that the restricted field strength
may be dominated by topological structures, which occur when one of
the parameters parametrising the colour field $n$ winds itself around
a non-analyticity in the colour field. If they exist, these objects
will lead to an area law scaling for the Wilson Loop and provide a
mechanism for quark confinement.  We search for these structures in
quenched lattice QCD. We perform the Abelian decomposition, and find
that the restricted field strength is dominated by peaks on the
lattice. Wilson Loops containing these peaks show a stronger area-Law
and thus provide the dominant contribution to the string tension.
\end{abstract}
\end{frontmatter}
\section{Introduction}

Colour confinement in QCD is one of the outstanding problems in physics. 
Although several possible confinement mechanisms have been proposed (for example, Abelian dominance \cite{thooft:1976,thooft:1981} 
or monopole condensation \cite{Nambu:1974,Monopoles1,Monopoles2,Mandelstam:1976,Polyakov:1977}), 
 none have been convincingly 
demonstrated to be correct. However, there has been important recent progress. Using a gauge invariant Abelian decomposition (the Cho-Duan-Ge (CDG) decomposition) and introducing the concept of the C-projection 
similar to the GSO-projection in string theory, Cho (and collaborators) have 
shown how to calculate the one-loop effective action of QCD gauge-invariantly 
and demonstrated that the effective potential condenses the monopole liquid \cite{Cho:2012pq}, implying that monopole condensation 
drives confinement. This project (of which this letter is the start) 
aims to verify and expand on this result in lattice QCD.

Lattice QCD has demonstrated the linear confining potential, but it has not 
been so successful determining what causes confinement. A popular mechanism 
studied in lattice QCD is Abelian dominance proposed by 'tHooft, which asserts 
that only the Abelian (i.e., neutral) part of QCD causes confinement \cite{Cho:1999ar}. 
This makes intuitive sense, since the coloured part of QCD  is confined. 
It has been studied on the lattice by decomposing the QCD potential into colour-neutral 
and coloured parts, using a gauge condition, such as the Maximal Abelian Gauge 
(MAG) or Laplacian Abelian Gauge to separate the Abelian part \cite{Kronfeld1987516,Suzuki:1989gp,Stack:1994wm,Shiba:1994db}. Similar approaches have been used to study monopole condensation in lattice QCD \cite{Arasaki:1996sm,Cea:2001an}.

This approach has serious defects. The whole
process is centred around fixing to one particular gauge, so it does
not demonstrate a gauge-invariant confinement mechanism. It also does not indicate what confines the colour. If an Abelian potential alone is enough for colour
confinement, we ought to have confinement in the Abelian QED.

However, the gauge invariant CDG decomposition
(also referred to as the Cho-Faddeev-Niemi decomposition)
avoids such defects
\cite{Cho:1980,Cho:1981,Duan:1979,F-N:98,Shabanov:1999}.  Unlike the
more popular MAG, this decomposition splits the QCD potential into
the restricted (neutral) part and the valence (coloured) part gauge
invariantly.  It also separates the topological part of the Abelian part of the gauge field. 
This decomposition can be used for a gauge-invariant
investigation of the topological basis of confinement.

Consider SU(2) QCD, and select a normalised ($n^an^a = 1$) Abelian direction $n\equiv \lambda^a n^a$, where $\lambda$ represents a Pauli matrix (or
Gell-Mann matrix in higher gauge groups). To construct the Abelian
decomposition, we impose the isometry condition on $A_\mu$ to
obtain the restricted potential $\hat{A}_\mu$
\begin{align}
D_\mu[\hatA] n=&0,& 
A_\mu^a \rightarrow \hat{A}^a_\mu=&{\cal B}_\mu^a+ {\cal C}_\mu^a, \nonumber\\
{\cal B}_\mu^a=&\frac{1}{2}n^a A^b_\mu n^b,& 
~~~{\cal C}_\mu^a=& \frac{i}{2g} \epsilon^{abc}n^b \partial_\mu n^c.
\label{eq:ic1}
\end{align}
The CDG decomposition arises by adding the valence potential $X_\mu=A_\mu-\hatA_\mu$
to the restricted potential
\begin{align}
A_\mu=& \hat A_\mu+ X_\mu={\cal B}_\mu+ {\cal C}_\mu+X_\mu, & \tr(n X_\mu)=&0. \label{eq:ic2}
\end{align}
The decomposition has several important features
\cite{Cho:1980,Cho:1981}: Firstly, the restricted potential, despite
being reduced, retains the full non-Abelian gauge degrees of
freedom. Secondly, the valence potential transforms gauge covariantly:
it represents the gauge covariant coloured gluons. Thirdly, the
decomposition is gauge invariant.  Once the Abelian direction $n$ is
chosen, the decomposition follows automatically regardless of the
choice of gauge.

But most importantly, the decomposition separates the topological potential gauge 
independently. The restricted potential $\hat A_\mu$ 
is made of two parts, the naive Abelian potential ${\cal B}_\mu$ and the topological 
potential ${\cal C}_\mu$, which can describe the Wu-Yang monopole~\cite{YangWu} when $n$ has an 
isolated point singularity representing the monopole topology 
$\pi_2(S^2)$ \cite{Cho:1980,Cho:1981}; this monopole structure is invariant 
under infinitesimal and analytic gauge transformations.

We aim (eventually) to identify the cause of confinement by examining the 
topological structures contained within a suitably chosen Abelian direction $n$ 
and their effects on the corresponding restricted field strength. The purpose 
of this initial letter is to demonstrate the feasibility of our approach, to isolate the topological potential 
and to suggest that it dominates 
the confining string. Later studies will further examine the consequences of this 
construction. We aim to confirm whether the topological structures in the restricted gauge field strength $\hat{F}_{\mu\nu}[\hatA]$ which, if they exist, might drive confinement can be associated with isolated monopoles, a condensed monopole/anti-monopole liquid, or some other topological structures.

To construct the Abelian decomposition on the
lattice, we must first choose the $N_C-1$ Abelian directions $n_j$, built from a $SU(N_C)$ matrix $\theta$, so $n_3 = \theta \lambda^3
\theta^\dagger$ ($N_C$ is the number of colours, the subscript $3$
indicates that $n_3$ is constructed from the third Gell-Mann matrix).
There are different ways of selecting $\theta$, including $\theta \in SU(N_C)/(SU(N_C-1)\times U(1))$ and $\theta \in
SU(N_C)/(U(1))^{N_C-1}$.  Choosing $\theta \in
SU(N_C)/(U(1))^{N_C-1}$ is advantageous as it contains all the possible
Abelian directions. It is important to select this
$\theta$ so all these configurations contribute to confinement
\cite{Cho:1980}.

In this letter we observe that we can always choose $\theta$ so 
that the static quark potential for the restricted field is identical to that of 
the full gauge field.\footnote{In practice, a different $n$ should be selected 
for each Wilson Loop to ensure that the restricted field can account for the 
confining potential. In this initial work, to save computer time, we use a single 
choice of $n$ for all our Wilson Loops, meaning that the link between the static potential of restricted and full QCD is inexact. 
Simulations without this simplification will 
be presented in a future work.} {\it We note that there always exists
a $SU(N_C)/(U(1))^{N_C-1}$ field $\theta$ which diagonalises the gauge links 
and removes the path ordering of the Wilson Loop, an observable used to measure the static potential.} By choosing the Abelian direction judiciously, we can always 
avoid the complicated path ordering in the Wilson loop and reduce it to an Abelian
form: we can always make the contribution of the valence potential $X_\mu$ to the 
Wilson loop vanish. This is natural: $X_\mu$ describes 
the coloured gluons and cannot play any role 
in confinement. This is Abelian dominance, which has been 
demonstrated theoretically \cite{Cho:1999ar}. But to show 
it by explicitly choosing a particular Abelian direction is really remarkable.  

Having thus selected $n$, we implement the isometry condition (\ref{eq:ic1}) on 
the lattice and construct the restricted field consistently, which allows us to 
express the Wilson Loop in terms of a surface integral over the restricted gauge 
field strength tensor. Our relationship for the string tension in terms of this 
restricted field is exact: we do not require any approximations or additional 
path integrals. We perform the lattice CDG decomposition, isolate the 
restricted potential $\hatA_\mu$ and the topological potential ${\cal C}_\mu$ 
and search for the topological structures in the restricted field strength, finding 
that they may cause an area law behaviour of the Wilson Loop. We outline how 
these topological structures arise in SU(2), leaving a fuller description and the 
extension to higher gauge groups to a subsequent work. If these structures exist
 (we do not prove in the theoretical analysis here that configurations containing 
them will contribute in practice) they will provide a mechanism for quark 
confinement.

By calculating the Wilson loop in a pure Yang Mills SU(3) lattice gauge theory with the full potential $A_\mu$, the 
restricted potential $\hatA_\mu$, and the topological potential ${\cal C}_\mu$, 
we pinpoint which potential generates the confining area law 
and is thus responsible for confinement. In this initial calculation, we concentrate 
on the string tension and an examination of the component of the restricted field 
responsible for confinement. Our result suggests that confinement is caused by the topological potential.

Similar lattice calculations, by the Chiba-KEK Lattice Group led by
Kondo \cite{Kondo:2005eq,Kondo:2010pt,Shibata:2007pi}, have recently
used the gauge independent Abelian decomposition to provide evidence for
monopole dominance in the confining potential. As far as we know,
these are the first gauge invariant lattice calculations to suggest
monopole dominance in the confining potential. The
most important difference between their work and ours is that they use
a different choice of $n$ whose $\theta$ is taken from a different
subgroup of $SU(3)$. These calculations, however, have unsatisfactory features. 
Firstly, their relationship between the Wilson Loop and the restricted 
field (based on~\cite{Diakonov:1989fc,Diakonov:1990uv}) requires a path 
integral over all possible $\theta$, in effect enlarging the gauge group by introducing a new $SU(3)/U(2)$ 
dynamical field. They then fixed $\theta$ (restoring the 
gauge group to SU(3)) by imposing the condition $[n,D^2 [A] n]=0$, which breaks the relationship between 
the Wilson Loops of the restricted and original gauge
fields. Secondly, they have chosen 
the `minimal' Abelian configuration for $n$ which leaves $SU(3)/(SU(2)\times U(1))$ 
invariant, choosing $n$ as only the $\lambda^8$-like 
Abelian direction, neglecting any contribution from the second Abelian direction 
constructed from $\lambda^3$. Clearly this $n$ can not describe the most general 
$SU(3)$ Abelian topologies. In this sense their monopole dominance is incomplete.

Here we construct the Abelian decomposition by 
rigorously imposing the isometry (\ref{eq:ic1}), choosing an Abelian direction $n$ which covers all possible 
SU(3) topological structures, and gives an exact equivalence between the Wilson Loops of the 
restricted and original gauge fields. 
This novel feature 
of our letter not only re-enforces the topological dominance but also makes 
it more precise. We will search for evidence of monopole condensation or some other mechanism in future work.

The letter is organized as follows. In section \ref{sec:2} we discuss the Abelian 
decomposition and its relation to the Wilson Loop and thus the static quark 
potential. In section \ref{sec:3} we discuss how topological structures which 
generate confinement may arise in this construction. We present numerical 
evidence in section \ref{sec:4}. We conclude in section \ref{sec:5}. Early results were presented in~\cite{Cundy:2012ee,lattice2013}.
 
\section{Abelian decomposition and Stokes' theorem}\label{sec:2}

We use the
convention that the superscript $a$ on a Gell-Mann matrix,
$\lambda^a$, implies that it should be summed over all values of $a$
($\lambda^aA^a \equiv\sum_{a = 1}^{N_C^2-1}\lambda^aA^a$), while the
index $j$ is restricted only to the diagonal Gell-Mann matrices (in
the standard representation
$A^j\lambda^j\equiv\sum_{j=3,8,\ldots,N_C^2-1}\lambda^jA^j$). The Wilson Loop, $W_L$, measures the confining potential in a theory with a $su(N_C)$ gauge field, $A_\mu=\frac{1}{2}A_\mu^a\lambda^a$~\cite{wilson:1977},
\begin{align}
W_L[C_s] = & \frac{ \tr\; W[C_s]}{N_C}, & W[C_s] = \mathcal{P}[e^{-ig\oint_{C_s} dx_\mu A_\mu(x)}].
\end{align}
$C_s$ is a closed curve of length $L$ which starts and finishes at a
position $s$ and $\mathcal{P}$ represents path ordering.   It is
expected that the vacuum expectation value of the Wilson Loop should
scale as $\langle W_L[C_s]\rangle\sim e^{-\rho\Sigma}$, where $\Sigma$
is the area of the surface enclosed by the curve $C_s$ and $\rho$ is
the string tension. We only consider planar Wilson Loops: $C_s$ is a
rectangle of temporal extent $T$ and spatial extent $R$ (we will later
restrict ourselves to loops in the $xt$ plane). The static quark
potential is given by $V(R)=-\lim_{T\rightarrow\infty}\log(\langle
W_L[C_s]\rangle)/T$. A linearly rising $V(R)$ is a signal for
confinement~\cite{wilson:1977}.

To define the path ordering, we split $C_s$ into infinitesimal segments of length $\delta\sigma$, with the gauge link $U_\sigma\in 
SU(N_C)=\mathcal{P}[e^{-ig\int_{\sigma}^{\sigma + \delta \sigma} A_\sigma d\sigma}]\sim e^{-ig\delta\sigma A_\sigma}$. $0\le\sigma\le L$ represents the position 
along the curve and $A_\sigma \equiv A_{\mu(\sigma)}(x(\sigma))$. We have assumed and will require throughout this work that the gauge field is 
differentiable.
This limits us to continuous gauge transformations (formed by repeatedly applying $A_\mu \rightarrow A_\mu + 
\frac{1}{g}\partial_\mu \alpha +i[A_\mu,\alpha]$ for infinitesimal and differentiable $\alpha\equiv\frac{1}{2}\alpha^a\lambda^a$). We also neglect the effects of the 
corners of the Wilson Loop (rounding them as necessary to avoid a discontinuity as $\sigma$ increases). 
This gives,
\begin{gather}
W[C_s]=\lim_{\delta\sigma\rightarrow0}\prod_{\sigma=0,\delta\sigma,2\delta\sigma,\ldots}^{L-\delta\sigma} U_\sigma.
\end{gather}
We introduce a field
$\theta_\sigma\equiv\theta(x(\sigma))\in U(N_C)$ and insert the
identity operator $\theta_\sigma \theta_\sigma^\dagger$ between each
pair of neighbouring gauge links on $C_s$. $\theta$ is chosen so that
$\theta^\dagger_\sigma U_\sigma\theta_{\sigma+\delta\sigma}$ is
diagonal.\footnote{The proof that this can be done for each link on a
  Wilson Loop is straight-forward, and shall be provided in the
  follow-up article.} $\theta_s$ therefore contains the eigenvectors of
$W[C_s]$ (the index $s$ indicates that $\theta_s$ refers to the field at the location where the Wilson Loop starts and ends).  As the phases of the eigenvectors are
arbitrary, this definition only determines $\theta$ up to a
$(U(1))^{N_C}$ transformation $\theta\rightarrow \theta\chi$. No physical observable depends on $\chi$, but in practice it is
useful to select the phases and ordering of the eigenvectors by some
arbitrary \textit{fixing condition}, giving a unique choice of
$\theta\in SU(N_C)/(U(1))^{N_C-1}$.  Under a gauge transformation,
$U_\sigma\rightarrow\Lambda_\sigma U_{\sigma}\Lambda_{\sigma
  +\delta\sigma}^\dagger$ for $\Lambda = e^{i\alpha^a \lambda^a}\in
SU(N_C)$, $\theta\rightarrow\Lambda\theta\chi$, where the
$(U(1))^{N_C-1}$ factor $\chi$ depends on the fixing condition. With
$\theta^\dagger_\sigma U_\sigma \theta_{\sigma + \delta\sigma} =
e^{i\sum_{\lambda^j \text{ diagonal}} \delta \sigma \hat{u}^j \lambda^j}$
for real $\hat{u}$,
\begin{gather}
\theta^\dagger_s W[C_s]\theta_s = e^{i\sum_{j = 3,8,\ldots} \lambda^j \oint_{C_s} d\sigma \hat{u}^j_\sigma},\label{eq:evth}
\end{gather}
removing the non-Abelian structure and the path ordering.

We will apply Stokes' theorem to express $W$ as a surface integral. First we extend the definition of $\theta$ and $\hat{u}^j$ across all space. For $\theta$, we 
construct nested curves in the same plane as $C_s$ and stack these curves on top of each other in the other dimensions. We define $\theta$ so it diagonalises 
each $W$ constructed from one of these curves. For $\hat{u}^j$, we construct a field $\hat{U}$ such that $\theta^\dagger(x) \hat{U}_\mu(x) 
\theta_{x+\delta\sigma\hat{\mu}}$ is diagonal $\forall x,\mu$ and $\hat{U}_\mu(x)=U_\mu (x)\forall x,\mu\in C_s$. Thus
\begin{align}
[\lambda^j,\theta_x^\dagger \hat{U}_{\mu,x}\theta_{x+\hat{\mu}\delta\sigma}] = &0, \\
\hat{U}_{\mu,x} n_{j,x+\delta\sigma\hat{\mu} }\hat{U}^\dagger_{\mu,x} - n_{j,x}= &0,&n_{j,x} \equiv &\theta_x \lambda^j \theta^\dagger_x\label{eq:defeq1}
\end{align}
are satisfied $\forall x,j$. Note that $n_j$ is independent of the choice of $\chi$. We relate $\hat{U}$ to the physical gauge field through a second field 
$\hat{X}$, defined by $U_\mu(x) = \hat{X}_\mu \hat{U}_\mu$. For later convenience (equation (\ref{eq:9})), we impose the condition
\begin{align}
\tr [n_{j,x}(\hat{X}^\dagger_{\mu,x} - \hat{X}_{\mu,x})] = &0
\label{eq:defeq2}.
\end{align}
We choose the solution to
equations (\ref{eq:defeq1}) and (\ref{eq:defeq2})
which maximises $\tr(\hat{X})$, a condition which is both gauge
invariant and satisfied along $C_s$ where $\hat{U}=U$ and $\hat{X} =
1$.
 Under a gauge transformation, $n_x \rightarrow \Lambda_x n_x
 \Lambda^\dagger_x$, $\hat{U}_\mu(x) \rightarrow \Lambda_x
 \hat{U}_{\mu,x} \Lambda^\dagger_{x+\hat{\mu}\delta\sigma}$ and
 $\hat{X}_{\mu,x} \rightarrow \Lambda_x \hat{X}_{\mu,x}
 \Lambda^\dagger_x$, so equations (\ref{eq:defeq1}) and
 (\ref{eq:defeq2}) are gauge-invariant.  Equations (\ref{eq:defeq1})
 and (\ref{eq:defeq2}) are lattice equivalents of the defining
 equations of the CDG
 decomposition~\cite{Cho:1980,Cho:1981,F-N:98,Shabanov:1999,Duan:1979},
 described in the continuum by the isometry condition (equations
 (\ref{eq:ic1}) and (\ref{eq:ic2}))
\begin{align}
&A_\mu = \hat{A}_\mu + X_\mu&
&D_\mu[\hat{A}] n_j =  0\nonumber\\
&D_\mu[\hat{A}] \alpha \equiv \partial_\mu \alpha - i g [\hat{A}_\mu,\alpha]& 
&\tr(n_j X_\mu)=0\nonumber\\
&\hat{A}_\mu = \frac{1}{2}\left[n_j\tr(n_j A_\mu) + \frac{i}{2g}  [n_j,\partial_\mu n_j]\right],
\end{align}
with $U_\mu \sim e^{-i\delta\sigma A_\mu}$ and $\hat{X}_\mu\sim
e^{-i\delta\sigma X_\mu}$ (to O($\delta\sigma^2$)).  

We express $\hat{U}$ as $\hat{U}_{\mu,x} \equiv \theta_{x}e^{i \lambda^j \delta\sigma\hat{u}^j_{\mu,x}} \theta^\dagger_{x+\hat{\mu}\delta\sigma}$ for 
real $\hat{u}$. Since $\hat{U}_\mu(x) = U_\mu(x) \forall x \in C_s$,  $W[C_s,U] = W[C_s,\hat{U}]=\theta_s W[C_s,\theta^\dagger 
\hat{U}\theta]\theta^\dagger_s = \theta_s e^{i \lambda^j \oint_{C_s} \hat{u}^j_\sigma d\sigma}\theta^\dagger_s$.
If $\hat{u}$ is differentiable, applying Stokes' theorem to this line integral gives
\begin{align}
\theta^\dagger_s W[C_s]\theta_s =& e^{i \lambda^j \int_{x \in \Sigma} d\Sigma_{\mu\nu} \hat{F}^j_{\mu\nu}},&
\hat{F}^j_{\mu\nu} =& \partial_\mu \hat{u}^j_\nu - \partial_\nu \hat{u}^j_\mu,\label{eq:7}
\end{align}
where $\hat{F}^j$ (like $\hat{u}$) is gauge invariant, $\Sigma$ the
(planar) surface bound by the curve $C_s$, and $d\Sigma$ an element of
area on that surface. Where $\hat{u}$ is not differentiable, we will have to break this integral into a surface integral over the region where $\hat{u}$ is analytic, and line integrals surrounding each of the non-analyticities in $\hat{u}$. We shall concentrate on the contribution from these non-analyticities below.

Through this choice of $\theta$, we have suggested that the dynamics
describing confinement can be expressed in terms of only an Abelian
field, and the suggestion and feasibility of using this choice of
$\theta$ as the basis of a CDG decomposition is the most important
novelty and result of this work. The coloured part of the gauge field,
$X_\mu$, does not contribute to confinement. We do not require any
additional path integrals. This procedure is gauge invariant, in the
sense that $\theta$ transforms gauge covariantly, and therefore the
restricted field strength $\hat{F}_{\mu\nu}$ and all other observables
constructed from the restricted field $\hat{A}$ are gauge invariant.

\section{Topological structures}\label{sec:3}

Now suppose that $\hat{u}^j$ contains a non-analyticity. We integrate
the field around a loop $\tilde{C}$ parametrised by $\tilde{\sigma}$
surrounding the discontinuity in $\hat{u}^j$, bounding the
surface integral by an additional line integral $ \oint_{\tilde{C}}
d\tilde{\sigma} \hat{u}^j_{\tilde{\sigma}}.  $ We define
$\{\tilde{C}_n\}$ as the set of curves surrounding all these
discontinuities, and $\tilde{\Sigma}$ the area bound within these
curves.  Thus
\begin{gather}
e^{i \lambda^j\delta\tilde{\sigma} \hat{u}^j_{\mu,x}} = \theta^\dagger_x \hat{X}^\dagger_{\mu,x} \theta_x \theta^\dagger_x U_{\mu,x} \theta_{x + 
\delta\tilde{\sigma}},\label{eq:str1}
\end{gather}
and $\hat{u}^j_{\mu,x}$ is continuous on $\tilde{C}$. After
gauge-fixing, we expand $U_\mu = 1 -i\frac{1}{2}g\delta\tilde{\sigma} A_\mu^a
\lambda^a$ and $\theta^\dagger_x \theta_{x+\delta\tilde{\sigma}} = 1 +
\delta\tilde{\sigma} \theta^\dagger_x \partial_{\tilde\sigma}
\theta_x$. We assume that $X_0\equiv
\frac{1}{2}\theta^\dagger_x(\hat{X}_{\mu,x} +
\hat{X}^\dagger_{\mu,x})\theta_x$, which will be close to the identity
operator, is well-defined
along $\tilde{C}$.
\begin{multline}
i\delta\tilde{\sigma} \hat{u}^j_{\mu,x} = \frac{1}{\tr (\lambda^j)^2}\text{Im}\left(\;\tr \left[\lambda^j\theta^\dagger_x \hat{X}^\dagger_{\mu,x} \theta_x 
\theta_x^\dagger U_{\mu,x}\theta_{x+\delta\tilde{\sigma} \hat{\mu}}\right]\right)\\=
\frac{1}{2\tr (\lambda^j)^2}\tr[\lambda^j \theta^\dagger_x (\hat{X}_{\mu,x}^\dagger - \hat{X}_{\mu,x}) \theta_x - \\\phantom{space}  i \lambda^j \delta \tilde{\sigma}\theta^\dagger_x [X_0]_{\mu,x} 
gA^a_{\mu,x} \lambda^a\theta_x + \\ 2\lambda^j \theta^\dagger [X_0]_{\mu,x} \theta_x \delta \tilde{\sigma}\theta_x^\dagger\partial_{\tilde{\sigma}} \theta]\label{eq:9}
\end{multline}
Using (\ref{eq:defeq2}), if $A_\mu$ and $X_0$ are analytic the final
term with a derivative in $\theta$ will dominate, giving
\begin{multline}
\theta^\dagger_s W[C_s]\theta_s = \exp\bigg(i  \lambda^j\bigg[ \int_{(x \in \Sigma) \cap (x\not{\in} \tilde{\Sigma})} d\Sigma_{\mu\nu} \hat{F}^j_{\mu\nu} +\\ 
{\sum_n}\oint_{\tilde{C}_n} d\tilde{\sigma}\frac{1}{\tr (\lambda^j)^2} \tr [\lambda^j X_0  \theta^\dagger \partial_{\tilde{\sigma}}\theta]\bigg]\bigg).
\end{multline}

There are three occasions when $\theta$ (and thus $\hat{u}$) may be discontinuous: if the Wilson Loop has degenerate eigenvalues; if the gauge field $A_\mu$ is discontinuous; but we will here concentrate on the possibility described below, which occurs in locations where $A_\mu$ is analytic
~\cite{CundyForthcoming}. 

In SU(2), we parametrise $\theta$ as
\begin{align}
\theta =&(\cos a \mathbb{I} + i \sin a \phi)e^{id_3\lambda^3}& \phi = & \left(\begin{array}{cc} 0&  e^{i c}\\ e^{-ic} &0 \end{array}\right)\nonumber\\
\bar{\phi} = &  \left(\begin{array}{cc} 0&  ie^{i c}\\ -ie^{-ic} &0 \end{array}\right)& \lambda^3 = & \left(\begin{array}{cc} 1& 0\\ 0 &- 
1\end{array}\right),\label{eq:thdec}
\end{align}
with $c \in \mathbb{R}$ and $0 \le a \le \pi/2$. $d_3 \in \mathbb{R}$ is determined by the fixing condition. $\mathbb{I}$ is the identity operator. As $\theta$ contains the 
eigenvectors of $W[C_s]$, it is differentiable except where $W[C_s]$ has degenerate eigenvalues and those points at $a = 0$ or $a \approx \pi/2$ where $c$ is 
ill-defined.  
We parametrise the plane of the Wilson Loop using polar coordinates $(r,\psi)$, with $r = 0$ at $a = \pi/2$.
At infinitesimal but non-zero $r$,  $c(r,\psi =0) = c(r,\psi = 2\pi) + 2\pi \nu_n$ for integer winding number $\nu_n$. With $c$ ill-defined at $r = 0$, we may 
find that $\nu_n \neq 0$. This will lead to the emergence of structures in $\hat{F}$ with a large field strength. $a$ and $c$ are not gauge invariant, 
so the corresponding structures in the gauge invariant $\hat{F}$ will be extended over a region rather than just a single point. This means that when we 
integrate the gauge invariant $\hat{u}$ along a curve around the singularity we should not choose the curve precisely at $a=\pi/2$, but some other path which both respects gauge invariance and contains the singularities in all gauges. We could, perhaps, use 
loops of non-vanishing magnetic current $k_\mu= \frac{1}{2} \epsilon_{\mu\nu\rho\sigma} \partial^{\nu} F^{\rho\sigma}$ to define this path (cf. ~\cite{Shibata:2009bg} and its 
references for a discussion of these loops), but, for simplicity, we have here assumed that we can construct a suitable curve at some constant $a=a_0$; while other choices might be better, they will not make any significant difference to our conclusions. 
It is straightforward to calculate
\begin{multline}
\theta^\dagger\partial_\sigma \theta = e^{-id_3\lambda^3}\big[ i \partial_\sigma a \phi + i \lambda^3 \partial_\sigma d +\\
 i \sin a \cos a \bar{\phi}\partial_\sigma c - i \sin^2 a \partial_\sigma c \lambda^3 \big]e^{id_3 \lambda^3}.
\end{multline}
We integrate along a path at fixed $a$ surrounding the
structure in $\hat{F}$, with a fixing condition holding $d_3$ constant,
\begin{align}
 \theta^\dagger_s  W[C_s]&\theta_s= e^{i \lambda^3\left[\int_{(x \in \Sigma) \cap (x\not{\in} \tilde{\Sigma})} d\Sigma_{\mu\nu} \hat{F}^j_{\mu\nu}- 
\sum_n\oint_{\tilde{C}_n} d\tilde{\sigma} \partial_{\tilde{\sigma}} c \sin^2 a_{0n}\right]} \nonumber\\
=& e^{i \lambda^3 \left[\int_{(x \in \Sigma) \cap (x\not{\in} \tilde{\Sigma})} d\Sigma_{\mu\nu} \hat{F}^j_{\mu\nu}-\sum_n 2\pi \nu_n \lambda^3\sin^2 
a_{0n}\right]}.
\end{align}
If $\nu_n \neq 0$ the structures arising from this discontinuity give
a significant contribution to the restricted field strength.  The
total Wilson Loop will be the product of a perimeter term, any remaining area law contribution from the surface integral over $\hat{F}_{xt}$, and
contributions from all these structures. As we can expect the number
of structures to be proportional to the area of the loop, this leads
to an area law for the Wilson Loop and a linear string
tension.

Although $\hat{F}$ (and therefore the structures) is gauge-invariant,
$\theta$, $\hat{X}$ and $\hat{U}$ depend on the gauge.  Since we
require that the gauge field is continuous, we can only use
continuous gauge transformations. However, to undo the winding in
$\theta$ requires a discontinuous gauge transformation.  Thus the
discontinuities in $\theta$ will survive any smooth gauge
transformation. For example, in SU(2), we can parametrise an
infinitesimal gauge transformation as
\begin{gather}
\Lambda = \left(\begin{array}{cc}\cos l_1& i \sin l_1 e^{il_2}\\ i \sin l_1 e^{-il_2} &\cos l_1\end{array}\right)\left(\begin{array}{cc} 
e^{il_3}&0\\0&e^{-il_3}\end{array}\right),
\end{gather}
with $l_1$ and $l_3$ infinitesimal and analytic and $0<l_2<2\pi$. If we fix $d_3 = 0$, we find that for $ |a| \gg O(l_1,l_3)$ and $|\pi/2-a| \gg O(l_1,l_3)$,
\begin{align}
c\rightarrow c' = &c+ 2l_3 + l_1 \sin(l_2-c)\cot a -l_1 \sin(l_2-c) \tan a \nonumber\\
a\rightarrow a' =& a +l_1 \frac{\cos(l_2-c)}{\cos a}
%
\end{align}
The winding number becomes 
\begin{gather}
\nu\rightarrow\oint \partial_{\tilde{\sigma}} c' d\tilde{\sigma} = 
\oint \partial_{\tilde{\sigma}} (c'-c) 
d\tilde{\sigma} + 2\pi \nu \,,
\end{gather}
and since $l_1$ and $l_3$ are infinitesimals and cannot change by
$2\pi$ and $(c'-c)$ is invariant under $c \rightarrow c + 2\pi \nu$,
the winding number is unaffected. The location where $a = \pi/2$ or $0$ may, however, be
shifted by a small amount.

In a SU($N_C$) gauge theory, we parametrise $\theta$ in terms of
$N_C-1$ diagonal elements $e^{id_j \lambda^j}$ and $(N_C^2 - N_C)/2$
matrices $e^{i a_i \phi_i} \in SU(2)/U(1)$, with each $\phi_i$ a
different embedding of equation (\ref{eq:thdec}) into
su($N_C$)~\cite{CundyForthcoming}. Since the different $\phi_i$ do not
commute, this parametrisation is not unique; nonetheless once the
parametrisation is fixed the analysis proceeds as in SU(2), and the
winding number is independent of the choice of parametrisation. There
will be a peak in $\hat{F}_{\mu\nu}$ whenever a $c_i$ winds around a
point where $a_i = \pi/2$ for any of the $SU(2)/U(1)$ matrices, and
each of those peaks contributes to the string tension.

\section{Numerical results}\label{sec:4}
We generated $16^3\times32$ and $20^3\times40$ quenched lattice QCD
(SU(3)) configurations with a Tadpole Improved Luscher-Weisz gauge
action~\cite{TILW} using a Hybrid Monte Carlo routine~\cite{HMC} (see
table \ref{tab:1}).  The lattice spacing was measured using the string
tension $\rho \sim (420 \text{MeV})^2$. We applied ten steps of
improved stout smearing~\cite{Morningstar:2003gk,Moran:2008ra} with
parameters $\rho_s = 0.015$ and $\epsilon = 0$. $\theta$ and $\hat{U}$
were extracted from the gauge field by solving equations
(\ref{eq:evth}), (\ref{eq:defeq1}) and (\ref{eq:defeq2})
numerically. Our algorithms and numerical set-up will be fully
described in a subsequent publication.
\begin{table}
{
\begin{center}
\begin{tabular}{|l l l l l|}
\hline
Lattice size & $L$ (fm)&$\beta$&$a$ (fm)& $\#$\\
\hline
$16^3 \times 32$& 2.30& 8.0&0.144(2) &91
\\
$16^3 \times 32$&1.84& 8.3 &0.114(1) &91
\\
$16^3 \times 32$&1.58 & 8.52 &0.099(1) &82
\\
$20^3 \times 40$&2.30 & 8.3&0.112(5) &61\\
\hline
\end{tabular}
\end{center}
}
\caption{Parameters for our simulations: the lattice size, the spatial
  extent of the lattice, $L$, the inverse gauge
  coupling $\beta$, the lattice spacing $a$, and the number of
  configurations in each ensemble $\#$. }\label{tab:1}
\end{table}

To match the continuum calculations, we need to work in
a continuous gauge, which is difficult
to realise on the lattice. We have therefore only used
gauge-invariant observables here, the string tension and restricted field strength. Extracting the components $a$ and $c$
from the gauge-dependent $\theta$ is straightforward: we presented some results for the
winding of $c$ around the peaks in~\cite{Cundy:2012ee}. However, it is
unclear what physical meaning can be given to this as we will be in a
different gauge to the continuum calculations. 

The string tension from the restricted potential is gauge invariant, and can be extracted from the Wilson 
Loop in a standard way. The restricted field strength can be measured using the gauge-invariant plaquette definition
\begin{multline}
e^{\hat{F}_{xt}^j(x+\frac{1}{2}a,t+\frac{1}{2}a) \lambda^j} \sim \\ \theta^\dagger_x \hat{U}_x(x,t) \hat{U}_t(x+a,t) 
\hat{U}^\dagger_x(x,t+a)\hat{U}^\dagger_t(x,t)  \theta_x,
\end{multline}
Constructing the $\theta$ contribution to the
restricted field strength is more challenging because the direct
calculation, measuring $\tr \lambda^j
(\theta^\dagger_x \theta_{x + a\hat{\mu}} - 1) \sim \tr \lambda^j
(\theta^\dagger_{x + \frac{1}{2}a\hat{\mu}}\partial_\mu \theta_{x +
  \frac{1}{2}a\hat{\mu}})$ for lattice spacing $a$, is not gauge-invariant.  A gauge
transformation which would be discontinuous in the continuum could
lead to additional discontinuities appearing in the observable or the
removal of discontinuities already present.  Fixing the gauge does not
help, as we might fix to a gauge where $A_\mu$ is discontinuous. We need to instead study the quantity
$\theta^\dagger_{x} \tilde{U}_{x,\mu} \theta_{x + \hat{\mu}}$ for some gauge
covariant field $\tilde{U}$ (so the whole expression is
gauge-invariant) which has only a minor effect on physical
observables such as the string tension so that only $\theta$
contributes to the Wilson Loop (the operator we use to represent $\tilde{U}$ is given later).

\begin{figure}
\begin{center}
\begin{tabular}{c}
{\tiny

\begingroup
  \makeatletter
  \providecommand\color[2][]{%
    \GenericError{(gnuplot) \space\space\space\@spaces}{%
      Package color not loaded in conjunction with
      terminal option `colourtext'%
    }{See the gnuplot documentation for explanation.%
    }{Either use 'blacktext' in gnuplot or load the package
      color.sty in LaTeX.}%
    \renewcommand\color[2][]{}%
  }%
  \providecommand\includegraphics[2][]{%
    \GenericError{(gnuplot) \space\space\space\@spaces}{%
      Package graphicx or graphics not loaded%
    }{See the gnuplot documentation for explanation.%
    }{The gnuplot epslatex terminal needs graphicx.sty or graphics.sty.}%
    \renewcommand\includegraphics[2][]{}%
  }%
  \providecommand\rotatebox[2]{#2}%
  \@ifundefined{ifGPcolor}{%
    \newif\ifGPcolor
    \GPcolorfalse
  }{}%
  \@ifundefined{ifGPblacktext}{%
    \newif\ifGPblacktext
    \GPblacktexttrue
  }{}%
  \let\gplgaddtomacro\g@addto@macro
  \gdef\gplbacktext{}%
  \gdef\gplfronttext{}%
  \makeatother
  \ifGPblacktext
    \def\colorrgb#1{}%
    \def\colorgray#1{}%
  \else
    \ifGPcolor
      \def\colorrgb#1{\color[rgb]{#1}}%
      \def\colorgray#1{\color[gray]{#1}}%
      \expandafter\def\csname LTw\endcsname{\color{white}}%
      \expandafter\def\csname LTb\endcsname{\color{black}}%
      \expandafter\def\csname LTa\endcsname{\color{black}}%
      \expandafter\def\csname LT0\endcsname{\color[rgb]{1,0,0}}%
      \expandafter\def\csname LT1\endcsname{\color[rgb]{0,1,0}}%
      \expandafter\def\csname LT2\endcsname{\color[rgb]{0,0,1}}%
      \expandafter\def\csname LT3\endcsname{\color[rgb]{1,0,1}}%
      \expandafter\def\csname LT4\endcsname{\color[rgb]{0,1,1}}%
      \expandafter\def\csname LT5\endcsname{\color[rgb]{1,1,0}}%
      \expandafter\def\csname LT6\endcsname{\color[rgb]{0,0,0}}%
      \expandafter\def\csname LT7\endcsname{\color[rgb]{1,0.3,0}}%
      \expandafter\def\csname LT8\endcsname{\color[rgb]{0.5,0.5,0.5}}%
    \else
      \def\colorrgb#1{\color{black}}%
      \def\colorgray#1{\color[gray]{#1}}%
      \expandafter\def\csname LTw\endcsname{\color{white}}%
      \expandafter\def\csname LTb\endcsname{\color{black}}%
      \expandafter\def\csname LTa\endcsname{\color{black}}%
      \expandafter\def\csname LT0\endcsname{\color{black}}%
      \expandafter\def\csname LT1\endcsname{\color{black}}%
      \expandafter\def\csname LT2\endcsname{\color{black}}%
      \expandafter\def\csname LT3\endcsname{\color{black}}%
      \expandafter\def\csname LT4\endcsname{\color{black}}%
      \expandafter\def\csname LT5\endcsname{\color{black}}%
      \expandafter\def\csname LT6\endcsname{\color{black}}%
      \expandafter\def\csname LT7\endcsname{\color{black}}%
      \expandafter\def\csname LT8\endcsname{\color{black}}%
    \fi
  \fi
  \setlength{\unitlength}{0.0500bp}%
  \begin{picture}(4818.00,2834.00)%
    \gplgaddtomacro\gplbacktext{%
      \csname LTb\endcsname%
      \put(860,640){\makebox(0,0)[r]{\strut{} 0}}%
      \put(860,817){\makebox(0,0)[r]{\strut{} 0.2}}%
      \put(860,994){\makebox(0,0)[r]{\strut{} 0.4}}%
      \put(860,1171){\makebox(0,0)[r]{\strut{} 0.6}}%
      \put(860,1348){\makebox(0,0)[r]{\strut{} 0.8}}%
      \put(860,1525){\makebox(0,0)[r]{\strut{} 1}}%
      \put(860,1702){\makebox(0,0)[r]{\strut{} 1.2}}%
      \put(860,1879){\makebox(0,0)[r]{\strut{} 1.4}}%
      \put(860,2056){\makebox(0,0)[r]{\strut{} 1.6}}%
      \put(860,2233){\makebox(0,0)[r]{\strut{} 1.8}}%
      \put(1502,440){\makebox(0,0){\strut{} 2}}%
      \put(2197,440){\makebox(0,0){\strut{} 4}}%
      \put(2892,440){\makebox(0,0){\strut{} 6}}%
      \put(3588,440){\makebox(0,0){\strut{} 8}}%
      \put(4283,440){\makebox(0,0){\strut{} 10}}%
      \put(160,1436){\rotatebox{-270}{\makebox(0,0){\strut{}$\log(\langle W[R,T]\rangle)/T$}}}%
      \put(2718,140){\makebox(0,0){\strut{}R}}%
      \put(2718,2533){\makebox(0,0){\strut{}$\beta = 8.52$}}%
    }%
    \gplgaddtomacro\gplfronttext{%
      \csname LTb\endcsname%
      \put(2457,2044){\makebox(0,0)[r]{\strut{}$U$ field $T = \infty$}}%
      \csname LTb\endcsname%
      \put(2457,1844){\makebox(0,0)[r]{\strut{}$\hat{U}$ field $T = \infty$}}%
      \csname LTb\endcsname%
      \put(2457,1644){\makebox(0,0)[r]{\strut{}$\tilde{U}_{1000}$ field $T = \infty$}}%
      \csname LTb\endcsname%
      \put(2457,1444){\makebox(0,0)[r]{\strut{}$\hat{\tilde{U}}_{1000}$ field $T = \infty$}}%
    }%
    \gplbacktext
    \put(0,0){\includegraphics{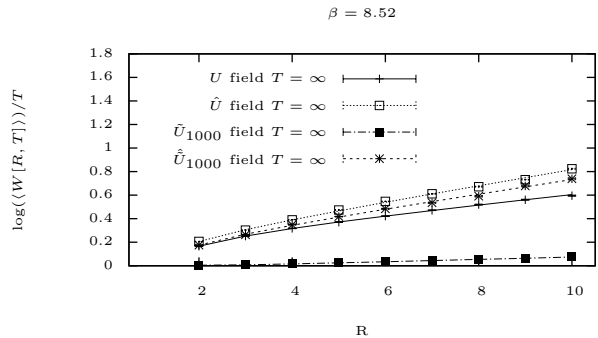}}%
    \gplfronttext
  \end{picture}%
\endgroup

}
\end{tabular}
\end{center}
\caption{The string tension extrapolated to infinite time for the original gauge field $U$, the restricted gauge field $\hat{U}$, the over-smeared field 
$\tilde{U}$ and the $\theta$ contribution to $\hat{U}$, $\hat{\tilde{U}}$, for a $\beta = 8.52$ ensemble. }\label{fig:5}
\end{figure}
 
Stout smearing~\cite{Morningstar:2003gk,Moran:2008ra} is a well-known tool to smooth the gauge field while preserving gauge invariance. Each Stout smearing sweep replaces $U_{x,\mu} \rightarrow 
U'_{x,\mu} = e^{iQ_x} U_{x,\mu}$ where $Q$ is a Hermitian operator constructed from closed loops of gauge links starting and finishing at $x$ (we constructed $Q$ from plaquettes and $2\times 1$ rectangles~\cite{Moran:2008ra}). A few smearing sweeps are often
used to remove unwanted discontinuous fluctuations in the gauge field. Too many smearing steps risk destroying the physical 
features of the gauge field. This is what we require: we set $\tilde{U}$ to be the gauge field $U$ subjected to a large number of stout smears: 
$\tilde{U}$ should resemble a pure gauge transformation, as any closed loop of gauge links will give the  identity operator and thus a zero field strength.
 We perform an Abelian decomposition on $\theta^\dagger_x \tilde{U}_{x,\mu} \theta_{x+\hat{\mu}}$ to extract the restricted field $\hat{ \tilde{U}}_{x,\mu}$
which satisfies $[\hat{\tilde{U}}_{x,\mu} ,\lambda^j] = 0$
and $\tr(\lambda^j(\tilde{X} -\tilde{X}^\dagger)) = 0$ with $\tilde{X} = \theta^\dagger_x \tilde{U}_{x,\mu} \theta_{x+\hat{\mu}} (\hat{ 
\tilde{U}}_{x,\mu})^{-1}$.
 We compare the field strength from this $\hat{\tilde{U}}$, representing the $\theta$ contribution to the restricted field strength, to the field strength from 
the restricted field $\hat{U}$.
We expect that the observables calculated from the $\theta$ field and the restricted field should be similar: the string tensions should be similar, and the field strengths should contain similar features.

\begin{table}
\begin{center}
{\begin{tabular}{|l|l l l l|}
\hline
$\beta$&8.0&8.3&8.52&8.3L\\
\hline
$U$&0.094(2)&0.064(3)&0.041(1)&0.059(1)
\\
$\hat{U}$&0.106(4)&0.087(2)&0.072(1)&0.095(1)
\\
$\tilde{U}_{100}$& 0.0835(4)&0.0536(3)&0.0413(3)&0.0554(3)
\\
$\hat{\tilde{U}}_{100}$ &0.111(5)  &0.080(2) & 0.071(2)&0.093(2)
\\
$\tilde{U}_{300}$& 0.0465(2)&0.0297(2)&0.0231(3)&0.0295(2)
\\
$\hat{\tilde{U}}_{300}$ &0.099(5)  &0.079(2) & 0.068(2)&0.091(2)
\\
$\tilde{U}_{500}$&0.0317(2) &0.0214(1)&0.0168(2)&0.0207(2)
\\
$\hat{\tilde{U}}_{500}$ & 0.096(5) & 0.080(2)& 0.067(2)&0.096(1)
\\
$\tilde{U}_{600}$& 0.0273(2)&0.0187(1)&0.0148(2)&0.0178(1)
\\
$\hat{\tilde{U}}_{600}$ &0.094(5)  &0.080(2) & 0.067(2)&0.093(1)
\\
$\tilde{U}_{800}$&0.0212(2) &0.0150(1) &0.0121(2) &0.0142(1)
\\
$\hat{\tilde{U}}_{800}$ &0.093(7)&0.080(2) &0.068(2)&0.092(2)
\\
$\tilde{U}_{1000}$&0.0173(2) &0.0123(1) &0.0103(2)&0.0119(1)
\\
$\hat{\tilde{U}}_{1000}$ &0.093(7) &0.080(2) &0.068(2)&0.092(2)
\\
\hline
$\frac{\hat{\tilde{U}}_{1000}}{\hat{U}}$&0.88(7)&0.92(3)&0.94(3)&0.97(2)
\\
$\frac{U}{\hat{U}}$&0.89(4)&0.74(4)&0.57(2)&0.62(1)
\\
\hline
\end{tabular}
}
\end{center}
\caption{The string tension extrapolated to infinite time  across all our ensembles. 8.3L refers to the $20^3\times40$ ensemble. The last two rows give the ratio of the topological and restricted string tensions and the restricted and actual string tensions.}\label{tab:4}
\end{table}

In figure \ref{fig:5} and table \ref{tab:4}, we extract the string
tension, $\rho$, for the original gauge field $U$, the restricted
field $\hat{U}$ and the $\theta$ contribution to the restricted field,
$\hat{\tilde{U}}$. We have calculated the expectation value of the $R\times T$
Wilson Loop in the $xt$ plane for one of the fields, and fit it to the function $\rho RT
+ aR + bT + c + dR/T + e T/R + f/T + g/R + h/(TR)$ for unknown
coefficients $\rho,a,\ldots h$. The cited errors are statistical, calculated using the
bootstrap method, and systematic, reflecting uncertainties in the fitting.  To reduce the computational overhead, for this
initial study we did not recalculate a new $\theta$ field for each
Wilson Loop but reused the same $\theta$ field for our whole
configuration, a simplification which destroys the identity between
the Wilson Loops for the $U$ and $\hat{U}$ fields; but is likely to keep any correlation between the $\hat{\tilde{U}}$ and $\hat{U}$ fields intact. We are currently
in the process of calculating the string tension with $\theta$
recalculated for each Wilson Loop, and intend to present the updated
result in a follow-up publication (early results are mentioned in~\cite{lattice2013}). We do not expect that the string tension between the $U$ and $\hat{U}$ fields will be identical, and, indeed, there is a large discrepancy in our results (which increases with decreasing lattice spacing). This would be particularly true for Wilson Loops not on the $xt$ plane (we have broken the hypercubic symmetry of the lattice by singling out the $xt$ plane while constructing $\theta$), so we have here restricted our study to Wilson Loops in the $xt$ plane. In this work, we are therefore more interested in the relationship between the string tension extracted from $\hat{\tilde{U}}$ and $\hat{U}$. Were these closely related, it would suggest that the topological ($\theta$) contribution to the string tension dominates, which is likely to be replicated in the full calculation where the string tensions for $\hat{U}$ and $U$ will be identical. 

\begin{figure}
\begin{center}
\begin{tabular}{c}
{\tiny
\begingroup
  \makeatletter
  \providecommand\color[2][]{%
    \GenericError{(gnuplot) \space\space\space\@spaces}{%
      Package color not loaded in conjunction with
      terminal option `colourtext'%
    }{See the gnuplot documentation for explanation.%
    }{Either use 'blacktext' in gnuplot or load the package
      color.sty in LaTeX.}%
    \renewcommand\color[2][]{}%
  }%
  \providecommand\includegraphics[2][]{%
    \GenericError{(gnuplot) \space\space\space\@spaces}{%
      Package graphicx or graphics not loaded%
    }{See the gnuplot documentation for explanation.%
    }{The gnuplot epslatex terminal needs graphicx.sty or graphics.sty.}%
    \renewcommand\includegraphics[2][]{}%
  }%
  \providecommand\rotatebox[2]{#2}%
  \@ifundefined{ifGPcolor}{%
    \newif\ifGPcolor
    \GPcolorfalse
  }{}%
  \@ifundefined{ifGPblacktext}{%
    \newif\ifGPblacktext
    \GPblacktexttrue
  }{}%
  \let\gplgaddtomacro\g@addto@macro
  \gdef\gplbacktext{}%
  \gdef\gplfronttext{}%
  \makeatother
  \ifGPblacktext
    \def\colorrgb#1{}%
    \def\colorgray#1{}%
  \else
    \ifGPcolor
      \def\colorrgb#1{\color[rgb]{#1}}%
      \def\colorgray#1{\color[gray]{#1}}%
      \expandafter\def\csname LTw\endcsname{\color{white}}%
      \expandafter\def\csname LTb\endcsname{\color{black}}%
      \expandafter\def\csname LTa\endcsname{\color{black}}%
      \expandafter\def\csname LT0\endcsname{\color[rgb]{1,0,0}}%
      \expandafter\def\csname LT1\endcsname{\color[rgb]{0,1,0}}%
      \expandafter\def\csname LT2\endcsname{\color[rgb]{0,0,1}}%
      \expandafter\def\csname LT3\endcsname{\color[rgb]{1,0,1}}%
      \expandafter\def\csname LT4\endcsname{\color[rgb]{0,1,1}}%
      \expandafter\def\csname LT5\endcsname{\color[rgb]{1,1,0}}%
      \expandafter\def\csname LT6\endcsname{\color[rgb]{0,0,0}}%
      \expandafter\def\csname LT7\endcsname{\color[rgb]{1,0.3,0}}%
      \expandafter\def\csname LT8\endcsname{\color[rgb]{0.5,0.5,0.5}}%
    \else
      \def\colorrgb#1{\color{black}}%
      \def\colorgray#1{\color[gray]{#1}}%
      \expandafter\def\csname LTw\endcsname{\color{white}}%
      \expandafter\def\csname LTb\endcsname{\color{black}}%
      \expandafter\def\csname LTa\endcsname{\color{black}}%
      \expandafter\def\csname LT0\endcsname{\color{black}}%
      \expandafter\def\csname LT1\endcsname{\color{black}}%
      \expandafter\def\csname LT2\endcsname{\color{black}}%
      \expandafter\def\csname LT3\endcsname{\color{black}}%
      \expandafter\def\csname LT4\endcsname{\color{black}}%
      \expandafter\def\csname LT5\endcsname{\color{black}}%
      \expandafter\def\csname LT6\endcsname{\color{black}}%
      \expandafter\def\csname LT7\endcsname{\color{black}}%
      \expandafter\def\csname LT8\endcsname{\color{black}}%
    \fi
  \fi
  \setlength{\unitlength}{0.0500bp}%
  \begin{picture}(5102.00,2380.00)%
    \gplgaddtomacro\gplbacktext{%
      \csname LTb\endcsname%
      \put(740,790){\makebox(0,0)[r]{\strut{} 4}}%
      \put(740,1090){\makebox(0,0)[r]{\strut{} 6}}%
      \put(740,1390){\makebox(0,0)[r]{\strut{} 8}}%
      \put(740,1689){\makebox(0,0)[r]{\strut{} 10}}%
      \put(740,1989){\makebox(0,0)[r]{\strut{} 12}}%
      \put(990,440){\makebox(0,0){\strut{} 5}}%
      \put(1640,440){\makebox(0,0){\strut{} 10}}%
      \put(2290,440){\makebox(0,0){\strut{} 15}}%
      \put(2940,440){\makebox(0,0){\strut{} 20}}%
      \put(3590,440){\makebox(0,0){\strut{} 25}}%
      \put(160,1389){\rotatebox{-270}{\makebox(0,0){\strut{}X}}}%
      \put(2420,140){\makebox(0,0){\strut{}T}}%
    }%
    \gplgaddtomacro\gplfronttext{%
      \csname LTb\endcsname%
      \put(4334,640){\makebox(0,0)[l]{\strut{}-2}}%
      \put(4334,827){\makebox(0,0)[l]{\strut{}-1.5}}%
      \put(4334,1014){\makebox(0,0)[l]{\strut{}-1}}%
      \put(4334,1202){\makebox(0,0)[l]{\strut{}-0.5}}%
      \put(4334,1389){\makebox(0,0)[l]{\strut{} 0}}%
      \put(4334,1576){\makebox(0,0)[l]{\strut{} 0.5}}%
      \put(4334,1764){\makebox(0,0)[l]{\strut{} 1}}%
      \put(4334,1951){\makebox(0,0)[l]{\strut{} 1.5}}%
      \put(4334,2139){\makebox(0,0)[l]{\strut{} 2}}%
    }%
    \gplbacktext
    \put(0,0){\includegraphics{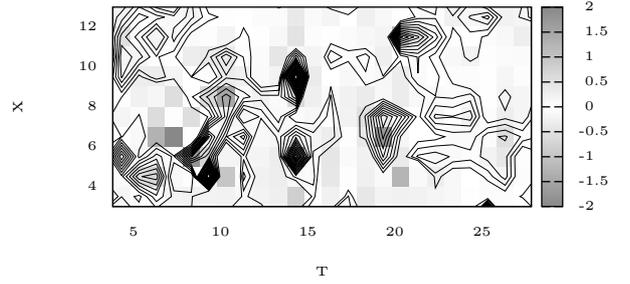}}%
    \gplfronttext
  \end{picture}%
\endgroup

}
\end{tabular}
\end{center}

\caption{A comparison between the peaks in $\hat{F}^{3}_{xt}$ (contours) against the topological field strength extracted from $\hat{ \tilde{U}}$ (shaded 
background). In this presentation the negative and positive peaks cannot be distinguished.  
We show one (typical) slice of the lattice at $Y=0$, $Z=11$.
 Due to the limited resolution of the lattice, the extrapolated contour lines and the shading have an error of up to one lattice spacing.}\label{fig:4}
\end{figure}

We calculate $\tilde{U}$ after 100, 300, 500, 600, 800 and 1000 sweeps of stout
smearing with parameters $\epsilon = 0, \rho_s = 0.1$
(following~\cite{Moran:2008ra}). We also show the string tension for
$\tilde{U}$, and can confirm that it is much smaller than that of the
original gauge field and decreases as we increase the level of
smearing. The string tension for $\hat{\tilde{U}}$ is unaffected by sufficiently large amounts of smearing,
suggesting that we have indeed measured the contribution from $\theta
\partial_\mu \theta^\dagger$ rather than any remnant of $U$ remaining after the smearing. 

 There is a considerable difference between the string tension for the original gauge field and for the restricted and topological fields, and this seems to increase as the lattice spacing decreases (the $\beta = 8.0$ $16^3$ ensemble and $20^3$ ensemble have roughly the same physical volume at different lattice spacings). The difference becomes very pronounced on the $\beta = 8.52$ ensemble. The $\hat{U}$ string tension has a weaker dependence on $\beta$ than that calculated from the actual gauge field. This is an artefact of the approximation we have made to accelerate the calculation, and both the approximation and its artefact will be removed in future work. Of more interest is the difference between the topological and restricted string tension, more likely to be duplicated in the calculation after our approximation has been removed. We see that the topological string tension appears to be slightly lower than the restricted tension on all our ensembles, by about $2\sigma$ or {88-97\%}. The variation of this discrepancy across our ensembles is not statistically significant. In all our ensembles the topological part of the restricted field dominates the restricted string tension.

Is the restricted field strength dominated by the
expected peaks? We use a contour plot to display the distribution of $\hat{F}^3_{xt}$ in figure
\ref{fig:4} on a slices of the lattice. The results for $\hat{F}^8_{xt}$ are similar. $\hat{F}^3_{xt}$ is indeed
dominated by objects one or two lattice spacings across. There is no correlation with the
structures on the neighbouring lattice slices, indicating that these
are point like objects rather than strings or surfaces. Do
these peaks emerge from the $\theta$ field? The background shading of
figure \ref{fig:4} shows the topological ($\hat{\tilde{U}}$) field strength, and there is a strong correspondence between the
location of the peaks in these two fields (albeit sometimes shifted by
a lattice spacing -- the resolution of our operators, and a few
structures visible in the topological field strength but not $\hat{F}_{xt}$). This pattern is repeated across all our ensembles.

\begin{figure}
\begin{center}
\begin{tabular}{c}
{\tiny
\begingroup
  \makeatletter
  \providecommand\color[2][]{%
    \GenericError{(gnuplot) \space\space\space\@spaces}{%
      Package color not loaded in conjunction with
      terminal option `colourtext'%
    }{See the gnuplot documentation for explanation.%
    }{Either use 'blacktext' in gnuplot or load the package
      color.sty in LaTeX.}%
    \renewcommand\color[2][]{}%
  }%
  \providecommand\includegraphics[2][]{%
    \GenericError{(gnuplot) \space\space\space\@spaces}{%
      Package graphicx or graphics not loaded%
    }{See the gnuplot documentation for explanation.%
    }{The gnuplot epslatex terminal needs graphicx.sty or graphics.sty.}%
    \renewcommand\includegraphics[2][]{}%
  }%
  \providecommand\rotatebox[2]{#2}%
  \@ifundefined{ifGPcolor}{%
    \newif\ifGPcolor
    \GPcolorfalse
  }{}%
  \@ifundefined{ifGPblacktext}{%
    \newif\ifGPblacktext
    \GPblacktexttrue
  }{}%
  \let\gplgaddtomacro\g@addto@macro
  \gdef\gplbacktext{}%
  \gdef\gplfronttext{}%
  \makeatother
  \ifGPblacktext
    \def\colorrgb#1{}%
    \def\colorgray#1{}%
  \else
    \ifGPcolor
      \def\colorrgb#1{\color[rgb]{#1}}%
      \def\colorgray#1{\color[gray]{#1}}%
      \expandafter\def\csname LTw\endcsname{\color{white}}%
      \expandafter\def\csname LTb\endcsname{\color{black}}%
      \expandafter\def\csname LTa\endcsname{\color{black}}%
      \expandafter\def\csname LT0\endcsname{\color[rgb]{1,0,0}}%
      \expandafter\def\csname LT1\endcsname{\color[rgb]{0,1,0}}%
      \expandafter\def\csname LT2\endcsname{\color[rgb]{0,0,1}}%
      \expandafter\def\csname LT3\endcsname{\color[rgb]{1,0,1}}%
      \expandafter\def\csname LT4\endcsname{\color[rgb]{0,1,1}}%
      \expandafter\def\csname LT5\endcsname{\color[rgb]{1,1,0}}%
      \expandafter\def\csname LT6\endcsname{\color[rgb]{0,0,0}}%
      \expandafter\def\csname LT7\endcsname{\color[rgb]{1,0.3,0}}%
      \expandafter\def\csname LT8\endcsname{\color[rgb]{0.5,0.5,0.5}}%
    \else
      \def\colorrgb#1{\color{black}}%
      \def\colorgray#1{\color[gray]{#1}}%
      \expandafter\def\csname LTw\endcsname{\color{white}}%
      \expandafter\def\csname LTb\endcsname{\color{black}}%
      \expandafter\def\csname LTa\endcsname{\color{black}}%
      \expandafter\def\csname LT0\endcsname{\color{black}}%
      \expandafter\def\csname LT1\endcsname{\color{black}}%
      \expandafter\def\csname LT2\endcsname{\color{black}}%
      \expandafter\def\csname LT3\endcsname{\color{black}}%
      \expandafter\def\csname LT4\endcsname{\color{black}}%
      \expandafter\def\csname LT5\endcsname{\color{black}}%
      \expandafter\def\csname LT6\endcsname{\color{black}}%
      \expandafter\def\csname LT7\endcsname{\color{black}}%
      \expandafter\def\csname LT8\endcsname{\color{black}}%
    \fi
  \fi
  \setlength{\unitlength}{0.0500bp}%
  \begin{picture}(4534.00,3118.00)%
    \gplgaddtomacro\gplbacktext{%
      \csname LTb\endcsname%
      \put(860,640){\makebox(0,0)[r]{\strut{} 0}}%
      \put(860,908){\makebox(0,0)[r]{\strut{} 0.2}}%
      \put(860,1176){\makebox(0,0)[r]{\strut{} 0.4}}%
      \put(860,1444){\makebox(0,0)[r]{\strut{} 0.6}}%
      \put(860,1713){\makebox(0,0)[r]{\strut{} 0.8}}%
      \put(860,1981){\makebox(0,0)[r]{\strut{} 1}}%
      \put(860,2249){\makebox(0,0)[r]{\strut{} 1.2}}%
      \put(860,2517){\makebox(0,0)[r]{\strut{} 1.4}}%
      \put(1379,440){\makebox(0,0){\strut{} 0}}%
      \put(1911,440){\makebox(0,0){\strut{} 2}}%
      \put(2443,440){\makebox(0,0){\strut{} 4}}%
      \put(2976,440){\makebox(0,0){\strut{} 6}}%
      \put(3508,440){\makebox(0,0){\strut{} 8}}%
      \put(4040,440){\makebox(0,0){\strut{} 10}}%
      \put(160,1578){\rotatebox{-270}{\makebox(0,0){\strut{}$\log(\langle W[R,T] \rangle)/T$}}}%
      \put(2576,140){\makebox(0,0){\strut{}R}}%
      \put(2576,2817){\makebox(0,0){\strut{}$\beta = 8.52$ Peaks in $\hat{F}_{xt}$ excluded, $T = \infty$}}%
    }%
    \gplgaddtomacro\gplfronttext{%
      \csname LTb\endcsname%
      \put(1847,2283){\makebox(0,0)[r]{\strut{}$\hat{U}$ field}}%
      \csname LTb\endcsname%
      \put(1847,2083){\makebox(0,0)[r]{\strut{}$\mathcal{K}=2.55$}}%
      \csname LTb\endcsname%
      \put(1847,1883){\makebox(0,0)[r]{\strut{}$\mathcal{K}=1.05$}}%
      \csname LTb\endcsname%
      \put(1847,1683){\makebox(0,0)[r]{\strut{}$\mathcal{K}=0.55$}}%
      \csname LTb\endcsname%
      \put(1847,1483){\makebox(0,0)[r]{\strut{}$\mathcal{K}=0.3$}}%
    }%
    \gplbacktext
    \put(0,0){\includegraphics{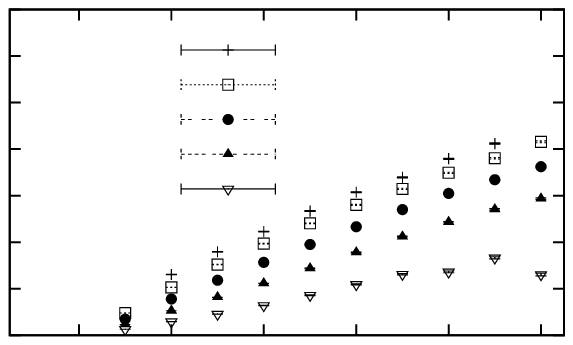}}%
    \gplfronttext
  \end{picture}%
\endgroup
}        

\end{tabular}
\end{center}
\caption{The $\hat{U}$ string tension, $\rho$, excluding Wilson loops
  containing peaks of height $|F_{xt}| > \mathcal{K}$ from the average
  ($\beta = 8.52$ ensemble). }\label{fig:1}
\end{figure}

We next investigate whether these peaks are responsible for the string
tension. Does excluding these peaks reduce or eliminate the
confining potential?  We usually measure the expectation value of the
Wilson Loop by averaging over every planar loop in the
configuration (in the $xt$ plane). Here we only include loops which do
not contain peaks higher than a cut-off $|\hat{F}| > \mathcal{K}$,
excluding those Wilson Loops which contain one of the peaks from the
average.  In figure \ref{fig:1} and table \ref{tab:6}, we see that the
string tension gradually decreases when averaging only over those
loops with $|F| < 1.0$ -- as expected if the peaks rather than the
fluctuations around zero are responsible for the confining
string. This pattern is again duplicated across our ensembles.

\begin{table}
\begin{center}
{
\begin {tabular}{|l|lll ll|}
\hline
$\mathcal{K}$&2.55&
1.30&
1.05&
0.55&
0.30
\\
\hline
$\beta 8.0$&12.1(5)&12.0(4)&11.5(3)&10.3(4)&5.00(1)
\\
$\beta 8.3$&9.2(1)&9.0(1)&8.8(1)&7.7(1)&3.82(1)
\\
$\beta 8.52$&7.7(1)&
7.8(1)&
7.6(1)&
6.7(1)&
5.0(3)
\\
$\beta 8.3L$&9.8(1)&9.0(1)&8.4(1)&7.3(1)&4.56(7)
\\
\hline
\end{tabular}
}
\end{center}
\caption{The $\hat{U}$ string tension, $\rho$ (in units of $10^{-2}a^{-1})$, excluding Wilson loops containing peaks of height $|F_{xt}| > \mathcal{K}$ 
from the average ($\beta = 8.52$ ensemble). $\beta 8.3L$ refers to the $20^340$ ensemble. }\label{tab:6}
\end{table}

\section{Conclusions}\label{sec:5}

We have proposed a method to express the Wilson Loop of an non-Abelian field 
in terms of an Abelian field without gauge fixing. Implementing the gauge
invariant Abelian decomposition (the CDG decomposition) on the lattice we 
relate the Wilson Loop to a surface integral over the CDG decomposition's restricted potential, 
and show that the restricted potential leads 
to an area law scaling for the quark-quark potential, and thus confinement. 
This confirms Abelian dominance of confinement.

The restricted potential contains two terms, one from the original 
gauge field (the naive Abelian part) and the other from the derivative of the 
$\theta$ field (the topological part). To isolate the cause of confinement, 
we must show which of these parts is most important for confinement. In this 
letter we have given evidence suggesting that the topological part dominates the 
Wilson Loop integral, and thus confines the colour. This strongly endorses 
the recent Chiba-KEK lattice calculations \cite{Kondo:2005eq,Kondo:2010pt}.

The Wilson loop describes the chromoelectric flux between quarks. 
While the topological part of the restricted gauge potential is 
known to contain coloured monopoles \cite{Cho:1980,Cho:1981}, so our lattice simulations are consistent with the recent theoretical analysis showing that monopole condensation generates confinement \cite{Cho:2012pq}, in this work we have only demonstrated that the topological 
potential is responsible for the area law of the Wilson loop. The structures we have found in the this component of the field strength are points rather than lines, suggesting that there is something else occurring (an isolated Wang-Yu monopole should have no contribution to the component of the field strength, $\hat{F}_{xt}$, studied here~\cite{CundyForthcoming}). The structures found in the electric field are certainly not isolated monopoles. More work is needed to compare our results with various models of the vacuum and see whether these objects are caused by monopole condensation or some other mechanism. 

Another important open question is how much the structures we have found in the restricted field will also be present in the restricted fields constructed from a different choice of $\theta$ (for example, by using a different set of nested Wilson Loops). If the structures are unique to this choice of $\theta$, then their use in identifying topological structures in the QCD vacuum causing confinement will be limited. However, since the gauge field strength $F_{\mu\nu}[A]$ can be decomposed as $F_{\mu\nu} = \hat{F}_{\mu\nu} + F[X]_{\mu\nu} - ig ([\hat{A}_\mu,X_\mu]- [\hat{A}_\nu,X_\nu])$, where $X_\mu$ does not contribute to confinement, we hypothesise that many structures found in $\hat{F}$ will be present in $F$; and that many structures leading to confinement contained within $F$ might be present in the $\hat{F}$ constructed from diverse choices of $\theta$. However, this hypothesis should be either confirmed or falsified in a future numerical study.   

Furthermore, we should also study the directional dependence of the field strength (and the Wilson Loop). In subsequent studies~\cite{lattice2013,CundyForthcoming}, we will consider the other components of the field strength tensor, finding that here the structures manifest themselves as one dimensional strings as well as points. In view of these later results, it is likely that the string tension will be larger if measured off the $xt$ plane. This discrepancy is only an artefact of our approximation, using the same $\theta$ for each Wilson Loop, and will be absent in a calculation without this simplification.

Our work is ongoing, and a full description of our theory and methods, 
and expanded numerical results, will be given in due course \cite{CundyForthcoming}.

\section*{Acknowledgments}
Numerical calculations used servers at Seoul National University
funded by the BK21 program of the NRF (MEST), Republic of
Korea. This research was supported by Basic Science Research Program through the National Research Foundation of Korea(NRF) funded by the Ministry of Education(2013057640). W. Lee is supported by the Creative Research Initiatives
Program (2013-003454) of the NRF grant, and acknowledges the support
from the KISTI supercomputing center through the strategic support
program for the supercomputing application research
(KSC-2012-G2-01). YMC is supported in part by the NRF grant
(2012-002-134) and by Konkuk University. We also thank the referee for several useful suggestions.


\begin{thebibliography}{10}
%
\expandafter\ifx\csname url\endcsname\relax
  \def\url#1{\texttt{#1}}\fi
\expandafter\ifx\csname urlprefix\endcsname\relax\def\urlprefix{URL }\fi
\expandafter\ifx\csname href\endcsname\relax
  \def\href#1#2{#2} \def\path#1{#1}\fi

\bibitem{thooft:1976}
G.~'t~Hooft, in: A.~Zichichi (Ed.), High Energy Physics, Editrice Comprostrini,
  Bologna, 1976.

\bibitem{thooft:1981}
G.~'t~Hooft, Nucl. Phys. B190 (1981) 455.

\bibitem{Nambu:1974}
Y.~Nambu, Phys. Rev. D10 (1974) 4262.

\bibitem{Monopoles1}
G.~'t~Hooft, Nucl. Phys. B79 (1974) 276.

\bibitem{Monopoles2}
A.~M. Polyakov, JETP Lett. 20 (1974) 194.

\bibitem{Mandelstam:1976}
S.~Mandelstam, Phys. Reports 23C (1976) 245.

\bibitem{Polyakov:1977}
A.~Polyakov, Nucl. Phys. B120 (1977) 429.

\bibitem{Cho:2012pq}
Y.~M. Cho, {Dimensional Transmutation by Monopole Condensation in QCD}, Phys.
  Rev. D87 (2013) 085025.
\newblock \href {http://arxiv.org/abs/1206.6936} {\path{arXiv:1206.6936}}.

\bibitem{Cho:1999ar}
Y.~M. Cho, {Abelian dominance in Wilson loops}, Phys. Rev. D62 (2000) 074009.
\newblock \href {http://arxiv.org/abs/hep-th/9905127}
  {\path{arXiv:hep-th/9905127}}, \href
  {http://dx.doi.org/10.1103/PhysRevD.62.074009}
  {\path{doi:10.1103/PhysRevD.62.074009}}.

\bibitem{Kronfeld1987516}
A.~Kronfeld, M.~Laursen, G.~Schierholz, U.-J. Wiese, Monopole condensation and
  color confinement, Phys. Lett. B 198~(4) (1987) 516 -- 520.
\newblock \href {http://dx.doi.org/10.1016/0370-2693(87)90910-5}
  {\path{doi:10.1016/0370-2693(87)90910-5}}.

\bibitem{Suzuki:1989gp}
T.~Suzuki, I.~Yotsuyanagi, {A possible evidence for Abelian dominance in quark
  confinement}, Phys. Rev. D42 (1990) 4257--4260.
\newblock \href {http://dx.doi.org/10.1103/PhysRevD.42.4257}
  {\path{doi:10.1103/PhysRevD.42.4257}}.

\bibitem{Stack:1994wm}
J.~D. Stack, S.~D. Neiman, R.~J. Wensley, {String tension from monopoles in
  SU(2) lattice gauge theory}, Phys. Rev. D50 (1994) 3399--3405.
\newblock \href {http://arxiv.org/abs/hep-lat/9404014}
  {\path{arXiv:hep-lat/9404014}}, \href
  {http://dx.doi.org/10.1103/PhysRevD.50.3399}
  {\path{doi:10.1103/PhysRevD.50.3399}}.

\bibitem{Shiba:1994db}
H.~Shiba, T.~Suzuki, {Monopole action and condensation in SU(2) QCD}, Phys.
  Lett. B351 (1995) 519--527.
\newblock \href {http://arxiv.org/abs/hep-lat/9408004}
  {\path{arXiv:hep-lat/9408004}}, \href
  {http://dx.doi.org/10.1016/0370-2693(95)00422-H}
  {\path{doi:10.1016/0370-2693(95)00422-H}}.

\bibitem{Arasaki:1996sm}
N.~Arasaki, S.~Ejiri, S.-i. Kitahara, Y.~Matsubara, T.~Suzuki, {Monopole action
  and monopole condensation in SU(3) lattice QCD}, Phys. Lett. B395 (1997)
  275--282.
\newblock \href {http://arxiv.org/abs/hep-lat/9608129}
  {\path{arXiv:hep-lat/9608129}}, \href
  {http://dx.doi.org/10.1016/S0370-2693(97)00066-X}
  {\path{doi:10.1016/S0370-2693(97)00066-X}}.

\bibitem{Cea:2001an}
P.~Cea, L.~Cosmai, {Abelian monopole and vortex condensation in lattice gauge
  theories}, JHEP 0111 (2001) 064.

\bibitem{Cho:1980}
Y.~M. Cho, Phys. Rev. D 21 (1980) 1080.

\bibitem{Cho:1981}
Y.~M. Cho, Phys. Rev. D 23 (1981) 2415.

\bibitem{Duan:1979}
Y.~Duan, M.~Ge, Sci. Sinica 11 (1979) 1072.

\bibitem{F-N:98}
L.~Faddeev, A.~Niemi, Phys. Rev. Lett. 82 (1999) 1624.
\newblock \href {http://arxiv.org/abs/hep-th/9807069}
  {\path{arXiv:hep-th/9807069}}.

\bibitem{Shabanov:1999}
S.~Shabanov, Phys. Lett. B 458 (1999) 322.
\newblock \href {http://arxiv.org/abs/hep-th/9903223}
  {\path{arXiv:hep-th/9903223}}.

\bibitem{YangWu}
T.~T. Wu, C.~N. Yang, in: H.~Mark, S.~Fernbach (Eds.), Properties of Matter
  under Unusual Conditions, Interscience, New York, 1969.

\bibitem{Kondo:2005eq}
K.-I. Kondo, T.~Murakami, T.~Shinohara, {Yang-Mills theory constructed from
  Cho-Faddeev-Niemi decomposition}, Prog.Theor.Phys. 115 (2006) 201--216.
\newblock \href {http://arxiv.org/abs/hep-th/0504107}
  {\path{arXiv:hep-th/0504107}}, \href {http://dx.doi.org/10.1143/PTP.115.201}
  {\path{doi:10.1143/PTP.115.201}}.

\bibitem{Kondo:2010pt}
K.-I. Kondo, A.~Shibata, T.~Shinohara, S.~Kato, {Non-Abelian Dual
  Superconductor Picture for Quark Confinement}, Phys. Rev. D83 (2011) 114016.
\newblock \href {http://arxiv.org/abs/1007.2696} {\path{arXiv:1007.2696}},
  \href {http://dx.doi.org/10.1103/PhysRevD.83.114016}
  {\path{doi:10.1103/PhysRevD.83.114016}}.

\bibitem{Shibata:2007pi}
A.~Shibata, et~al., {Toward gauge independent study of confinement in SU(3)
  Yang-Mills theory}, POS LATTICE-2007 (2007) 331.
\newblock \href {http://arxiv.org/abs/0710.3221} {\path{arXiv:0710.3221}}.

\bibitem{Diakonov:1989fc}
D.~Diakonov, V.~Y. Petrov, {A FORMULA FOR THE WILSON LOOP}, Phys. Lett. B224
  (1989) 131--135.
\newblock \href {http://dx.doi.org/10.1016/0370-2693(89)91062-9}
  {\path{doi:10.1016/0370-2693(89)91062-9}}.

\bibitem{Diakonov:1990uv}
D.~Diakonov, V.~Y. Petrov, {Generating functional for Yang-Mills theory and the
  confinement requirement}, Phys. Lett. B242 (1990) 425--428.
\newblock \href {http://dx.doi.org/10.1016/0370-2693(90)91787-C}
  {\path{doi:10.1016/0370-2693(90)91787-C}}.

\bibitem{Cundy:2012ee}
N.~Cundy, W.~Lee, J.~Leem, Y.~Cho, {String tension from gauge invariant
  Magnetic Monopoles}, PoS LATTICE2012 (2012) 213.
\newblock \href {http://arxiv.org/abs/1211.0664} {\path{arXiv:1211.0664}}.

\bibitem{lattice2013}
N.~Cundy, Y.~Cho, W.~Lee, The gauge invariant abelian decomposition, {PoS}
  {(LATTICE2013)} (2013) 471.
\newblock \href {http://arxiv.org/abs/1311.3029} {\path{arXiv:1311.3029}}.

\bibitem{wilson:1977}
K.~G. Wilson, in: M.~Levy, P.~Mitter (Eds.), New Developments in Quantum Field
  Theory and Statistical Mechanics, Plenum, New York, 1977.

\bibitem{Shibata:2009bg}
A.~Shibata, et~al., {Topological configurations of Yang-Mills field responsible
  for magnetic-monopole loops as quark confiner}, PoS LAT2009 (2009) 232.
\newblock \href {http://arxiv.org/abs/0911.4533} {\path{arXiv:0911.4533}}.

\bibitem{TILW}
M.~L{\"u}scher, P.~Weisz, Commun Math Phys 97 (1985) 59.

\bibitem{HMC}
S.~Duane, A.~Kennedy, B.~Pendleton, D.~Roweth, Phys. Lett. B195 (1987) 216.

\bibitem{Morningstar:2003gk}
C.~Morningstar, M.~J. Peardon, Phys. Rev. D69 (2004) 054501.
\newblock \href {http://arxiv.org/abs/hep-lat/0311018}
  {\path{arXiv:hep-lat/0311018}}.

\bibitem{Moran:2008ra}
P.~J. Moran, D.~B. Leinweber, Phys. Rev. D77 (2008) 094501.
\newblock \href {http://arxiv.org/abs/0801.1165} {\path{arXiv:0801.1165}},
  \href {http://dx.doi.org/10.1103/PhysRevD.77.094501}
  {\path{doi:10.1103/PhysRevD.77.094501}}.

\bibitem{CundyForthcoming}
N.~Cundy, Y.~M. Cho, W.~Lee, in preparation.


 \end{thebibliography}
\end{document}